\documentclass[groupedaddress,superscriptaddress,
reprint,
amsmath,amssymb,
aps,
colorlinks=true, breaklinks=true, linkcolor=darkblue, citecolor=darkblue, urlcolor = darkblue
]{revtex4-2}
\usepackage{cancel}
\usepackage{graphicx}
\usepackage{dcolumn}
\usepackage{bm}
\usepackage{stmaryrd}

\usepackage{siunitx}
\usepackage{orcidlink}

\usepackage{times}

\usepackage{xcolor}
\definecolor{darkblue}{rgb}{0, 0, 0.8}
\usepackage{lipsum}
\newcommand{\PASQAL}{Pasqal SAS, 24 rue Emile Baudot, 91120 Palaiseau, Paris, France}
\newcommand{\IOGS}{Université Paris-Saclay, Institut d’Optique Graduate School,
CNRS, Laboratoire Charles Fabry, 91127 Palaiseau Cedex, France}

\begin{document}


\title{Defect-free arrays at the thousand-atom scale in a 4-K cryogenic environment}
\author{Desiree~Lim}
\thanks{These authors contributed equally to this work.}
\affiliation{\PASQAL}
\author{Hadriel~Mamann} 
\thanks{These authors contributed equally to this work.}
\affiliation{\PASQAL}
\author{Grégoire~Pichard}
\thanks{These authors contributed equally to this work.}
\affiliation{\PASQAL}\affiliation{\IOGS}
\author{Lilian~Bourachot} 
\affiliation{\PASQAL}
\author{Arvid~Lindberg}
\affiliation{\PASQAL}
\author{Clotilde~Hamot}
\affiliation{\PASQAL}
\author{Hugo~Le~Bars}
\affiliation{\PASQAL}
\author{Florian~Fasola}
\affiliation{\PASQAL}
\author{Siddhy~Tan}
\affiliation{\PASQAL}
\date{\today}
\author{Gwennolé~Cournez}
\affiliation{\PASQAL}
\author{Sylvain~Dutartre}
\affiliation{\PASQAL}
\author{Thierry~Cartry}
\affiliation{\PASQAL}
\author{Sylvain~Lemettre}
\affiliation{\PASQAL}
\author{Richard~Hostein}
\affiliation{\PASQAL}
\author{Julien~Paris}
\affiliation{\PASQAL}
\author{Franck~Ferreyrol}
\affiliation{\PASQAL}
\author{Andréa~Collardey}
\affiliation{\PASQAL}
\author{Adrien~Signoles}
\affiliation{\PASQAL}
\author{Thierry~Lahaye}
\affiliation{\IOGS}
\author{Corentin~Monmeyran}
\thanks{Corresponding author: \href{mailto:corentin.monmeyran@pasqal.com}{corentin.monmeyran@pasqal.com}}
\affiliation{\PASQAL}
\author{Bruno~Ximenez}
\affiliation{\PASQAL}

\begin{abstract}
We report on a cryogenic platform at 4~K incorporating high numerical aperture optics for the generation of large-scale tweezers arrays, and compatible with Rydberg-state manipulation. We achieve trapping lifetimes of around 5000~s, significantly extending the available experimental time for the preparation of large-scale arrays. By combining two trapping lasers at different wavelengths and by minimizing other atom losses during the rearrangement and imaging processes, we demonstrate the preparation of defect-free arrays with up to 1024 atoms. Our cryogenic design opens exciting prospects for analog and digital quantum computing. 
\end{abstract}

\maketitle


\section{Introduction}

Neutral atom arrays have rapidly emerged as a powerful platform for quantum engineering. Over the past decade, arrays of individually trapped neutral atoms in optical tweezers have proven to be a versatile platform for both analog quantum simulation~\cite{Bernien2017,deLeseleuc2019,Omran2019,Ebadi2021, Scholl2021, Chen2023,Chen2025}  and digital quantum computation~\cite{Levine2018, Levine2019,Bluvstein2022,Evered2023,Tsai2025}. Advances in coherent control, laser stabilization, and system engineering have enabled high-performance neutral-atom quantum processors with high gate fidelities~\cite{Levine2018, Levine2019, Evered2023}, reliable state preparation and measurement, and increasingly scalable architectures. Combining these features with continuous-loading techniques will enable much higher repetition rates~\cite{Chiu2025, Li2025}.

One of the key challenges in scaling neutral atom platforms is the preparation of large registers. During this process, atoms can be lost due to several mechanisms: collisions with background gas, imaging losses, and losses due to atom transport. However, when scaling systems to thousands or tens of thousands of atoms, vacuum-limited losses become the dominant factor, making long atomic lifetimes essential. Operating the system in a cryogenic environment is one of the easiest ways to attain extreme-high vacuum (XHV), as the surfaces of the system become efficient cryopumps. Atom lifetimes of up to 6000~s has been demonstrated under such conditions~\cite{Schymik2021}. Besides allowing for larger registers, another advantage of using a cryogenic system is that it reduces the effects of blackbody radiation (BBR). Indeed, at room temperature, BBR drives transitions between neighboring Rydberg states, shortening the effective Rydberg-state lifetime and thereby limiting the fidelity of Rydberg-mediated operations. For low angular momentum Rydberg levels, removing the contribution of BBR by working at low temperatures increases the Rydberg state lifetime by a factor two to three. For \emph{circular} Rydberg states, which currently attract a growing interest for quantum computing and quantum simulation \cite{Cohen2021,Ravon2023,Wu2023,Hoelzl2024}, the gain in lifetime can reach several orders of magnitude~\cite{Nguyen2018}.

Recently, several groups have demonstrated the feasibility of preparing large neutral-atom arrays in cryogenic environments. A key design consideration for this is to maintain high optical access in such systems while keeping an effective shielding of the cryogenic region from room-temperature radiation~\cite{Schymik2022,Pichard2024, Zhang2025, Jin2026}. In our previous work, the use of high numerical aperture objectives enabled the preparation of arrays containing more than 800 atoms~\cite{Pichard2024}. However, insufficient shielding between the room-temperature and cryogenic regions limited the achievable atomic lifetimes to at most a few hundred seconds, preventing the achievement of defect-free arrays. With proper shielding, it was shown in~\cite{Zhang2025} that it is possible to combine high optical access with significantly improved atomic lifetimes of up to 3000~s and Rydberg excitation. Very recently, Jin \emph{et al.} demonstrated a significant increase of the lifetime  of low angular momentum Rydberg states in a cryogenic environment~\cite{Jin2026}. However, these results were obtained in tweezers arrays that are still of moderate sizes. 

In this work, we realize a cryogenic platform that accommodates in-vacuum high numerical aperture objectives outside cryogenic shields at 4~K and 30~K that are sealed with windows. With this design, we measure vacuum-limited lifetimes of around 5000~s that can be sustained over long periods of time before degrading, and that can be swiftly recovered with a fast regeneration method. In order to generate large-scale optical tweezers arrays, we combine two trapping lasers at different wavelengths that form independent trap arrays. Leveraging extended atom lifetimes alongside with good imaging and atom transport performance, we demonstrate the assembly of large atom arrays of up to 1024 atoms, with a very high probability of obtaining defect-free registers, in excess of 10\,\%, and an average defect rate of 0.3\,\%. 

\vfill

\section{Experimental setup} 

\begin{figure*}[tb!]
    \centering
    \includegraphics[]{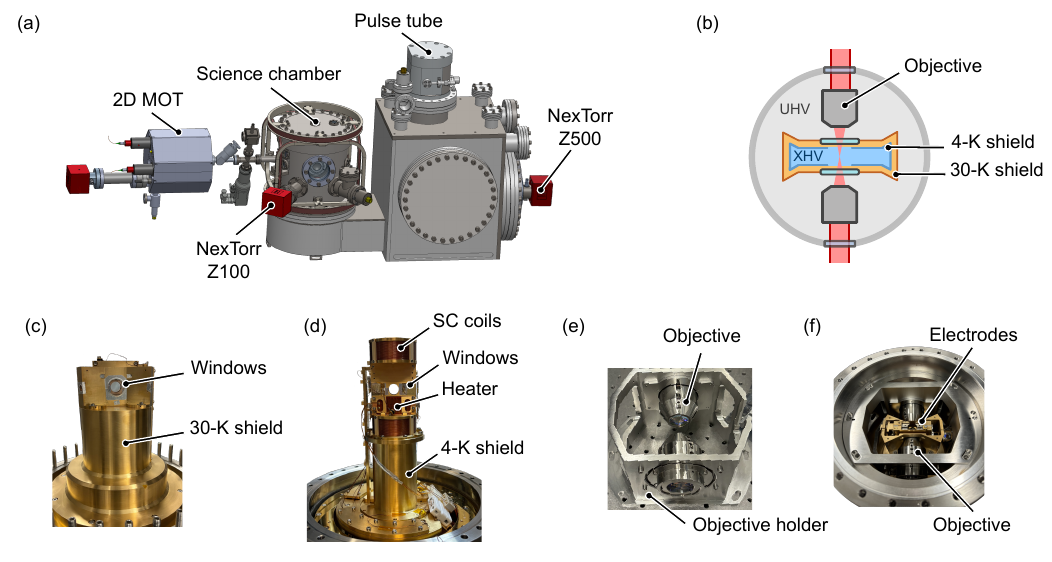}
    \caption{Overview of the experimental apparatus. (a): CAD rendering of the vacuum system. From left to right, the atomic source, the science chamber containing the objectives, and the technical chamber accommodating the pulse tube refrigerator. (b): Schematic diagram showing the incorporation of the objectives with the cryogenic shields. The 300-K region is in the ultra-high vacuum (UHV) regime, in contrast with the extreme-high vacuum (XHV) regime inside the 4-K shield. (c): The outer, 30-K shield, with windows attached on the apertures using aluminum tape. (d): The inner, 4-K shield, partially opened to show the superconducting (SC) coils and windows on the MOT beam ports. (e):~The stainless steel microscope objective holder at 300~K, after the pre-alignment of the objectives. (f): Assembly of the objectives and shields in the science vacuum chamber. Electrodes are placed on the 4-K shield for compensating stray electric fields in future experiments with Rydberg states.}
    \label{fig:experimentalSetup}
\end{figure*}

\subsection{Cryogenic system and optical core}

\paragraph*{Design considerations.---}
Several requirements need to be met to achieve good vacuum performance in a cryogenic environment within a vacuum chamber. A key limitation to the vacuum quality near the atomic plane is the flux of gas particles from the 300-K ultra-high vacuum (UHV) region entering the cold, XHV, region where the atoms are located. Over time, this flux contaminates the cryogenic surfaces, thereby reducing their cryopumping efficiency. This issue can be mitigated in several ways. First, we can minimize the overall outgassing of the vacuum components, for example by treating their surfaces. Second, the pumping speed of the 300-K region can also be increased. Third, the conductance between the room-temperature region and the cold region can be reduced, for example by limiting the size and number of apertures connecting the two regions. Finally, we can also enhance the effective pumping speed of the cryogenic stage, for instance by enlarging the available cryogenic surface area.


The design of the current setup is an upgrade from our previous implementation~\cite{Pichard2024}. In that system, residual gas analysis (RGA) suggested that hydrogen outgassing from the stainless-steel walls of the room temperature vacuum chamber was the dominant source of background gas in the 4-K region. To mitigate this effect in the new setup, we use a science chamber that was vacuum-fired prior to installation. In addition, optical access to the atomic plane is now provided through a series of windows attached on the thermal shields. These windows interrupt direct line-of-sight paths between the 300-K environment and the cold region while preserving the optical access required for trapping and imaging. Having windows on the optical path of the microscope objectives comes at a cost: the objectives need to be designed taking the windows into account, and the position and tilt of the windows need to be precisely controlled to avoid optical aberrations. For this reason, we chose to install a window only on the outer, 30-K shield, and leave the aperture on the inner, 4-K shield windowless.   

\begin{figure*}[t]
    \centering
    \includegraphics[]{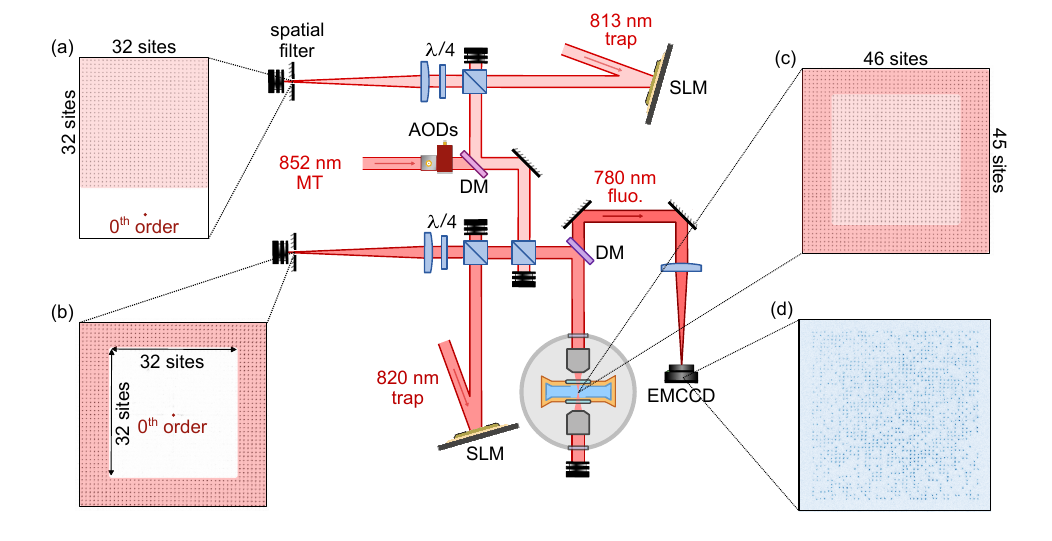}
    \caption{Schematic diagram of the optical layout for
    generating large arrays by combining two trapping lasers. 
     Each trapping laser beam (at 813~nm and 820~nm) is reflected off a separate SLM, and then spatially filtered in the Fourier plane to remove the zeroth diffraction order using a mirror with a hole followed by a beam dump. The two beams are then combined on a polarizing beam-splitter (PBS) before being sent into the science chamber with a dichroic mirror (DM). Before this, the 852~nm moving tweezers (MT) beam is combined with the 813~nm path, also with a dichroic mirror. (a) The inner part of the array, with the zeroth order on the side. (b) The outer part of the array, with the zeroth order in the center. (c) The resulting $45\times46$ tweezers array seen by the atoms. (d) Fluorescence image of $1170$ single atoms randomly loaded in the array.}
    \label{fig:opticalsetup}
\end{figure*}

\paragraph*{Cryogenic system.---}
The overall apparatus is depicted in Fig.~\ref{fig:experimentalSetup}~(a). Compared to our previous work~\cite{Pichard2024}, we still use a 2D-MOT as the atomic source and a two-stage pulse-tube refrigerator for cryo-cooling; however the science chamber was entirely re-designed to accommodate the custom-made objectives (Special Optics, $\mathrm{NA}=0.6$, with a 14~mm working distance and $\pm250\;\mu$m field of view). A NEG-ion combined pump (NexTorr Z100, with a pumping speed of $\sim150$~L$/$s for hydrogen) is connected to the science chamber, while a larger model (NexTorr Z500, offering a higher pumping speed of $\sim580$~L$/$s for hydrogen) is mounted on the technical chamber.

To allow the various lasers and the atomic beam to reach the trapping region, apertures are  made on the cryogenic shields as shown in Fig.~\ref{fig:experimentalSetup}~(b) and Fig.~\ref{fig:experimentalSetup}~(c). All of the apertures for lasers, except the one of the trapping beam on the $4\text{-}\mathrm{K}$ shield, are sealed with AR-coated fused-silica windows. The window faces in direct view of the atoms are coated with a 40-nm thick Indium-Tin-Oxide (ITO) layer to avoid the possible accumulation of stray electric charges. These  windows are centered on their respective apertures using a shallow counter-bore and are secured on the shields using aluminum tape, in order to ensure correct thermalization and low stress levels (we did not observe any degradation in the polarization state of the MOT beams through the windows after cool-down, ruling out any stressed-induced birefringence in the windows). In this design, the atomic beam aperture which is mandatory for allowing Rb atoms to enter the chamber is the only direct line of sight to the atomic plane. This aperture is kept small with an area of only $\sim0.4\,\mathrm{cm^2}$. The use of windows on almost all apertures is a crucial improvement for the vacuum quality compared to the previous version of our setup~\cite{Pichard2024}. 

Two coils are mounted above and below the objectives on the $4\text{-}\mathrm{K}$ shield as shown in Fig.~\ref{fig:experimentalSetup}~(d). The superconducting wires used for the coils ensure negligible heat load from the current. The coils were designed to provide a magnetic field gradient of $\sim10\,\mathrm{G/cm/A}$ in the anti-Helmholtz configuration and a bias field of $\sim 20\,\mathrm{G/A}$ in the Helmholtz configuration at the atomic plane. Vertical slits are cut in the $30\text{-}\mathrm{K}$ and $4\text{-}\mathrm{K}$ shields to prevent the formation of eddy currents when the magnetic fields are varied during an experimental sequence. These slits are left open but were designed so as to avoid direct line-of-sight from the room-temperature region to the atomic plane. In addition, eight electrodes placed around the objectives allow for the compensation of stray electric fields in  future experiments with Rydberg states.

\paragraph*{Optical core.---}
The optical core of the experimental apparatus lies in the center of the science chamber, in which optical tweezers are generated by focusing the trap beam with the microscope objective. At the same position, three retro-reflected beams (with a $1/e^2$ diameter of 4.8~mm) that enter via the MOT apertures in the shields intersect to create a 3D MOT. The objectives are mounted on a stainless-steel objective holder (Fig.~\ref{fig:experimentalSetup}~(e)) and are pre-aligned at $300\,\mathrm{K}$. The holder is then carefully mounted on the top of the science chamber. A reference mirror, placed above the entrance pupil of the objective, is used for aligning beams into the optical axis. With their field of view of $\pm250\,\mu\mathrm{m}$, the objective allows for the generation of large arrays, comprising in principle several thousands of tweezers. 

\subsection{Generation of large tweezers arrays}

The optical setup for generating the trap arrays is shown in Fig.~\ref{fig:opticalsetup}. Two fiber amplifier systems are used for trap generation: one at $813\,\mathrm{nm}$ and one at $820\,\mathrm{nm}$ (Precilasers, models FL-SF-813-10-CW and FL-SF-820-10-CW), each with an output power of about $6\,\mathrm{W}$. To maximize the total power available for trapping, we combine the lasers by sending each beam to a separate spatial light modulator (SLM, Hamamatsu, X15213-02), before combining them with a polarizing beam-splitter (PBS). 

A major source of power losses for the traps comes from the finite diffraction efficiency of the SLM, which significantly decreases at large diffraction angles. This motivates choosing trap patterns centered around the zeroth diffraction order. The main issue with this approach however is the centered residual zeroth order, which distorts traps near the center of the pattern and makes atom detection difficult. We therefore spatially filter the zeroth order using the approach of  Ref.~\cite{Manetsch2025}. Each trapping beam is sent to a separate SLM before being focused in a Fourier plane where we place a custom mirror with a $500\text{-}\mu\mathrm{m}$ diameter hole. The two beams are then later combined with a PBS before entering the science chamber. We chose to combine the two lasers with a PBS to avoid using a dichroic mirror designed for two wavelengths very close to one another. With the $820\text{-}\mathrm{nm}$ beam, we generated a centered array of $46\times45$ traps with a spacing of $4.5\,\mu\mathrm{m}$, including a square ``hole'' of $32\times32$ sites at the center. The $813\text{-}\mathrm{nm}$ beam was used to generate a $32\times32$ array that was translated away from the zeroth order to fill this central region. While centering the array over the zeroth order would require about $27\,\%$ less power, translating the pattern allows us to fill the hole in the center of the trap layout. Figure~\ref{fig:opticalsetup}~(c) and (d) show, respectively, the result of aligning the two patterns into a combined array of 2070~traps, and a fluorescence image of single Rb atoms trapped in the combined array.

On the $813\text{-}\mathrm{nm}$ path, an $852\text{-}\mathrm{nm}$ beam, sent through a pair of orthogonal acousto-optic deflectors (AODs, AA Opto Electronic, DTSXY-400-850), serves as a moving tweezers for rearrangement. The moving tweezers is combined with the $813\text{-}\mathrm{nm}$ beam via a dichroic mirror (DI02-R830-50.8-D, Semrock), before the two trap beams are combined. The AODs are controlled by a custom FPGA board driving fast DACs~\cite{Pichard2024}.

\begin{figure}[tbh!]
    \centering
    \includegraphics[width=\columnwidth]{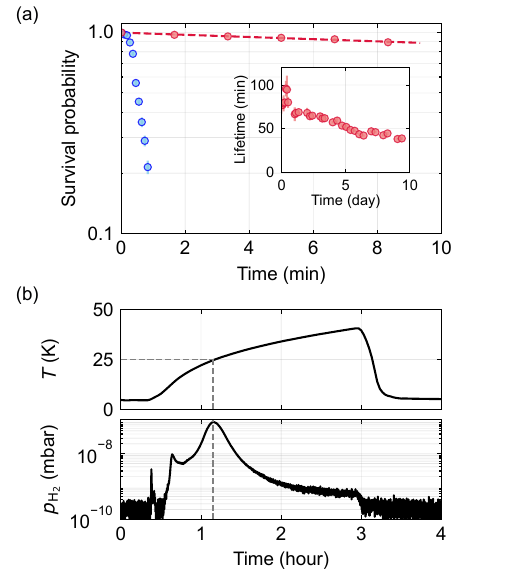}
    \caption{(a) Measurement of the trapping lifetime performed just after a fast regeneration event (see text). Blue symbols: without polarization gradient cooling (PGC) pulses; the heating of atoms in the traps leads to a short ($<1$~min) lifetime. Red symbols: with PGC pulses, a fit by an exponential decay (red dashed line) gives a $1/e$ lifetime of $\sim 80$~min. Inset: evolution of the measured lifetime (with PGC pulses) following the regeneration event. (b) During a fast regeneration, the $4\text{-}\mathrm{K}$ shield warms up to $\sim 40\,\mathrm{K}$, allowing to degas the hydrogen molecules that were cryosorbed on the shield. A large hydrogen peak is visible on the RGA when the shield reaches a temperature of $\sim 25\,\mathrm{K}$ (grey dashed line).}
    \label{fig:lifetime}
\end{figure}

\subsection{Lifetime in the new cryogenic system}

At the start of the experiment, atoms are loaded into the tweezers array with the 3D MOT, for which we use three pairs of retro-reflected beams red-detuned by $2.9\,\Gamma$ from $F=2\to F'=3$ transition of the $\mathrm{D}_2$ line. For the measurement of in-tweezers atom lifetime, we typically use an array of $204$ traps with a $5\text{-}\mathrm{\mu m}$ spacing at $820\,\mathrm{nm}$, corresponding to around $450\,\mathrm{mW}$ in the atomic plane. We apply pulsed polarization gradient cooling (PGC) beams detuned by $-6.6\,\Gamma$ from the $F=2\to F'=3$ bare transition for $2\,\mathrm{ms}$ every second to mitigate heating of the atoms in the traps. This allows to accurately measure the vacuum-limited lifetime of the trapped atoms~\cite{Schymik2021}. 

In Fig.~\ref{fig:lifetime}~(a), we measure a trapping lifetime of 80 minutes indicating extreme-high vacuum inside the 4~K region. However, this measured lifetime degrades over the course of a few days, reaching 40 minutes after 10 days as shown in the inset, and about 20 minutes after 45 days (not shown). This is due to the deterioration of the vacuum quality over time as the cryogenic shields gradually become saturated. As hydrogen, the main residual gas in the chamber, is only pumped efficiently at 4 K through cryosorption~\cite{Baglin2020}, an increase in surface coverage directly leads to a reduction of the pumping efficiency~\cite{Chill2019}. However, it is possible to quickly recover the vacuum performance through a method that we term ``fast regeneration'', that we now describe; it was performed prior to the measurements shown in Fig.~\ref{fig:lifetime}~(a). 

To restore the vacuum performance, one needs to degas the particles cryosorbed by the thermal shields and pump them out of the system. A first possibility (``full regeneration'') that allows us to entirely recover the vacuum performance is to turn off the  cryostat and let the entire science chamber warm up to room temperature. As the temperature increases, the particles trapped on the thermal shields are degassed. Turbo-molecular pumps (Pfeiffer, HiCube 80) evacuate the residual gases. We also regenerate the getters in the NexTorr pumps by heating them while keeping the turbomolecular pumps attached. After this full regeneration process, the next cool-down begins with clean surfaces in the chamber, allowing to restore the vacuum performance. This process is however quite time consuming, as the system takes around four days to reach room temperature. This is why we use a significantly quicker process, fast regeneration, that restores most of the vacuum performance in just a few hours. Instead of degassing all the particles on the thermal shields, the cryostat is only turned off for approximately $3$ hours allowing the $4\text{-}\mathrm{K}$ shield to reach only around $40\,\mathrm{K}$. As shown in Fig.~\ref{fig:lifetime}~(b), we observe on the RGA that a huge peak of hydrogen is detected near $25\,\mathrm{K}$, which indicates that a large quantity of hydrogen is desorbed at this moment. The molecules are then pumped by the attached ion pumps and getters. In future work, we will entirely automate the regeneration process with dedicated heaters that are already installed on the 4~K shield, as shown in Fig.~\ref{fig:experimentalSetup}~(d).

\begin{figure}[tb!]
    \centering
    \includegraphics[width=\columnwidth]{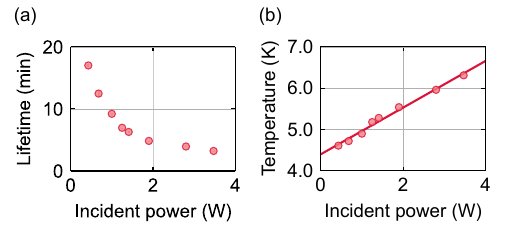}
    \caption{(a) Atom lifetime measurement as a function of the total incident power sent into the chamber. These measurements were performed 45 days after a fast regeneration had been performed. (b) Temperature of the 4-K shield as a function of the incident optical power. }
\label{fig:thermal_effects}
\end{figure}

The above lifetime measurements were performed in medium-size tweezer  arrays. However, in order to realize large-scale arrays, one needs to send several watts of laser trap power into the chamber. It is thus important to assess how the trapping lifetime depends on the incident optical power, and thus on the heat load seen by the cryostat. At high trap powers, we measure significantly lower lifetimes, as can be seen in Fig.~\ref{fig:thermal_effects}~(a). For instance, 45 days after the fast regeneration (at a stage where the cryogenic shields were approaching saturation), we measured trapping lifetimes as long as 18~min at low powers, but of only 3~min at a trap power of 4~W sent into the system. This is likely due to the desorption of residual gases when the temperatures of the shields increase due to the heat load. To confirm this interpretation, we measured the temperature of the 4-K shield as a function of the total incident power sent into the chamber, and find an increase of about $0.6~\text{K}/\text{W}$ from fitting the data in Fig.~\ref{fig:thermal_effects}~(b).  

\section{Rearrangement of large arrays}

\begin{figure*}[tb!]
    \centering
    \includegraphics[]{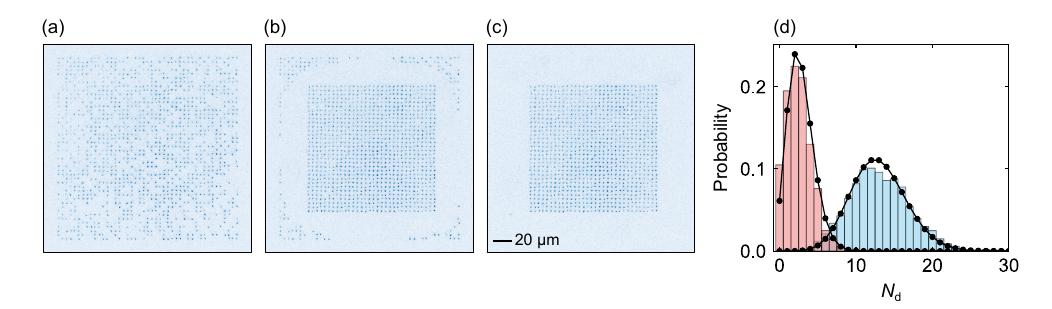}
    \caption{Rearrangement of a 1024-atom defect-free array. Single-shot fluorescence images (a) after atoms are loaded in an array comprising 2070 traps, (b) after a first cycle of rearrangement (note the remaining reservoir atoms at the periphery and the few holes in the target array) and (c) after two cycles, with the resulting defect-free array of 1024~atoms. (d) Distribution of the number $N_{\rm d}$ of defects after a single cycle (blue bars) or two cycles (red bars) of rearrangement. The prediction of the simple model described in Appendix~\ref{sec:appendix} is shown by the black curves with the success probability of moves $p_{\mathrm{move}}=99.05 \,\%$ and the survival probability of fixed atoms $p_{\mathrm{survival}}=99.75 \,\%$ measured independently. The mean number of moves is 1014 during the first rearrangement step and decreases to 135 during the second step.  } 
    \label{fig:Rearrangement_1024}
\end{figure*}

To illustrate the added value of long trapping lifetimes, we perform rearrangement of atoms into a 1024-trap target array, starting from a layout of 2070 traps with a depth of around $1\,\mathrm{mK}$ each. The first step of the rearrangement is to acquire an image and evaluate where the atoms have been loaded. Based on this first image, an algorithm (the LSAP2 algorithm from Ref.~\cite{Schymik2020}) is used to calculate the list of moves to fill the register. It minimizes the number of moves and avoids collisions between atoms. In addition, we perform multiple cycles of rearrangement to compensate some defects introduced during the first rearrangement cycle. The number of defects in the register after the second cycle drops for two main reasons. First, fewer moves are required, resulting in fewer losses due to atom transport. Moreover, the rearrangement time is shorter and hence losses from background-gas collisions are reduced. Histograms of the distribution of the number $N_{\rm d}$ of defects after one and two rearrangement cycles are presented in~Fig.~\ref{fig:Rearrangement_1024}. Using the two cycles, we obtain defect-free, 1024-atom arrays in more than 10\,\% of the experimental realization, and an average defect rate of 0.3\,\%, hinting at very low probabilities of (i) background-gas collisions, (ii) imaging losses, and (iii) atom transport losses. In the following, we estimate the probabilities of these processes from the mean number $\langle N_{\rm d} \rangle$ of defects. We then show that a simple model, assuming that those losses are uncorrelated, gives a full distribution of $N_{\rm d}$ which is in good agreement with experimental data.

Collisions with the background gas can eject atoms from their traps. The corresponding loss probability is given by $p_{\mathrm{vac-loss}} = 1 - \exp(-\Delta t/\tau) $, where $\Delta t$ is the rearrangement time and $\tau \simeq 2000 $~s is the measured lifetime (lower than the maximum lifetime achieved due to shield heating induced by laser power). In our experiment, $\Delta t$ corresponds to the time interval between the images acquired before and after rearrangement, namely $2\,\mathrm{s}$ for the first cycle and $1.5\,\mathrm{s}$ for the second cycle. This yields vacuum-induced loss probabilities of $p_{\mathrm{vac-loss}} = (0.10 \pm 0.01)\,\%$ and $(0.08 \pm 0.01)\,\%$, respectively.

The second source of defects originates from detection-induced losses, as atoms can be expelled from the traps by the imaging light. The probability associated with this mechanism is $p_{\mathrm{det-loss}}=(0.15 \pm 0.03)\,\%$. This number was obtained by measuring atom losses between two successive images with no operations in between and subtracting losses due to collisions with the background gas.

The third contribution is related to the rearrangement process itself, specifically the transport operations performed by the moving tweezers from initial sites to targeted sites in the register area. Atoms moved during rearrangement have a probability $p_\mathrm{t}=p_{\mathrm{move}}~p_{\mathrm{survival}}=(98.70 \pm 0.02)\,\%$ to be successfully transferred, with $p_{\mathrm{survival}}=(1-p_{\mathrm{vac-loss}})(1 -p_{\mathrm{det-loss}})=(99.75 \pm 0.03)\,\%$ the survival probability of fixed atoms and $p_{\mathrm{move}}=(99.05 \pm 0.05)\,\%$ the success probability of moves. Experimentally, the value of $p_\mathrm{t}$ was obtained by measuring the fraction of atoms loaded into their target sites after each rearrangement step. The value of $p_{\mathrm{move}}$ was then inferred from $p_{\mathrm{t}}$ and $p_{\mathrm{survival}}$. 

The solid lines shown in Fig.~\ref{fig:Rearrangement_1024} correspond to the predictions of our theoretical model on the defect distribution of our register, which incorporate the different loss probabilities described above as input parameters (see Appendix). The model reproduces the experimental data well, indicating that the dominant physical processes are correctly captured. However, a discrepancy is observed for the defect-free probability after the second step, which is higher experimentally than predicted ($10.5\,\%$ instead of $6\,\%$). Since the model relies on averaged parameters and assumes homogeneous move losses across the $2070$ traps, we believe that residual spatial inhomogeneities in transport fidelity account for this deviation. 

\section{Conclusion and outlook}

In conclusion, we developed a new cryogenic setup enabling the generation of large-scale tweezers array with very long trapping lifetimes around 5000~s. We implemented an optical setup based on two SLMs capable of generating more than 2000 optical traps. Using this platform, we demonstrated the rearrangement of 1024 atoms with an average defect rate of 0.3\,\%, implying that defect-free registers are obtained in about 10\,\% of the experiments. Larger atom arrays should be in principle achievable without major changes in the setup, as our current limitation arises mainly from the available laser power. 

One of the challenges ahead for tweezers array platforms is to increase experimental repetition rates. A possible solution is to implement continuous-loading protocols, in which the MOT is loaded in a separate chamber and used to refill reservoir traps while rearrangement and Rydberg excitation sequences are performed simultaneously~\cite{Chiu2025,Li2025}. Integrating the cryogenic setup with these protocols will require some design efforts.

Another important future step will be to perform Rydberg excitation in this setup and demonstrate a suppression of BBR-induced transitions, as seen recently in Ref.~\cite{Jin2026}. For quantum simulation, either with low-angular momentum states (not only for the case of direct Rydberg excitation but also for Rydberg dressing experiments~\cite{Jau2016,Goldschmidt2016,Hollerith2022,Steinert2023}), or when using circular Rydberg states, this will enable longer coherence times. For quantum computing, extended Rydberg lifetimes should directly result in improved gate fidelities, while a better vacuum will reduce atom loss and correspondingly the overhead of error correction. 

\appendix
\section{A simple model for the distribution of the number of defects in a rearranged array}
\label{sec:appendix}

We describe here the model used in Fig.~\ref{fig:Rearrangement_1024} to account for the distribution of the number of defects in a final rearranged array. We distinguish between the loss of atoms that are transported during rearrangement and loss of atoms that remain fixed. We model the first contribution by a random variable $\mathcal{T}$ (for transported) and the second contribution by a random variable $\mathcal{F}$ (for fixed). $\mathbb{P}(\mathcal{T}=k)$ is the probability to get $k$ defects for atoms that have been moved and $\mathbb{P}(\mathcal{F}=k)$, the probability of having $k$ defects for fixed atoms. These are given by binomial distributions:
\begin{align}
    \mathcal{T} &\sim \mathrm{Binomial}(N_{\mathrm{m}}, 1-p_{\mathrm{t}})\\
    \mathcal{F} &\sim \mathrm{Binomial}(N_{\mathrm{r}}-N_{\mathrm{m}}, 1-p_{\mathrm{survival}})
\end{align}
where $N_{\mathrm{r}}$ is the register size, $N_{\mathrm{m}}$ the number of moves, and $p_{\mathrm{survival}}$ is the survival probability of fixed atoms during rearrangement. It can be decomposed in two loss mechanisms, detection losses $p_{\mathrm{det-loss}}$ and losses due to background-gas collisions $p_{\mathrm{vac-loss}} = 1 - \exp(-\Delta t/\tau) $, where $\Delta t$ is the rearrangement time and $\tau$ the vacuum-limited lifetime. We can therefore write the survival probability of fixed atoms as $p_{\mathrm{survival}}=(1-p_{\mathrm{vac-loss}})(1 -p_{\mathrm{det-loss}})$. The transfer move probability is $p_{\mathrm{t}}=p_{\mathrm{move}}~p_{\mathrm{survival}}$ with $p_{\mathrm{move}}$ the move efficiency corrected by the losses of fixed atoms (detection and vacuum-induced losses).

With this we can estimate the probability of having $N_\mathrm{d}$ defects in the final register:
\begin{equation}
    \mathbb{P}(N_\mathrm{d}) = \sum_{k\in I} \mathbb{P}(\mathcal{T}=k)\,\mathbb{P}(\mathcal{F}=N_\mathrm{d}-k).
\end{equation}
We can rewrite this equation as:
\begin{widetext}
\begin{equation}
\mathbb{P}(N_\mathrm{d}) = \sum_{k\in I}\binom{N_{\mathrm{m}}}{k}p_{\mathrm{t}}^{N_{\mathrm{m}}-k}(1-p_{\mathrm{t}})^k\binom{N_{\mathrm{r}}-N_{\mathrm{m}}}{N_\mathrm{d}-k}(1-p_{\mathrm{survival}})^{N_\mathrm{d}-k}p_{\mathrm{survival}}^{N_{\mathrm{r}}-N_{\mathrm{m}}-(N_\mathrm{d}-k)},
 \label{number_defects}
\end{equation}
\end{widetext}
with $I=\llbracket \mathrm{max}(0,N_\mathrm{d}+N_{\mathrm{m}}-N_{\mathrm{r}}), \mathrm{min}(N_{\mathrm{m}},N_\mathrm{d}) \rrbracket$ in order to keep positive exponents. The mean defect $\langle N_{\mathrm{d}} \rangle$
after one rearrangement cycle is obtained by calculating the sum 
$\sum N_{\mathrm{d}} \, P(N_{\mathrm{d}})$
with Eq.~(\ref{number_defects}) and the set of parameters given above.
We observed experimentally that, for a compact layout, the number of moves to perform in the first cycle is approximately equal to the size of the register ($N_{\mathrm{m}}\approx N_{\mathrm{r}}$).
For the second cycle, $N_{\mathrm{m}}$ is lower than the first one and is proportional to the number of defects to fill. Since we consider a compact register of $N \times N$ atoms, each defect after the first step requires a large number of moves to be corrected. In the worst case, the defects are located in the central region of the register and 
$N_{\mathrm{m}}/{N_{\mathrm{d}}} = N/2 $.
On average, 
$\langle N_{\mathrm{m}}/{N_{\mathrm{d}}}\rangle \approx N/4,$ 
which corresponds to defects located at the midpoint between the center and the edge of the register.

\hfill
\begin{acknowledgments}
We thank Davide Dreon for fruitful discussions at the early stage of this project, and the entire Pasqal team for the support that made this work possible. We also thank the Vacuum Surfaces and Coatings Group at CERN for the advice on the design of the vacuum system, and the CERN Quantum Technology Initiative,  which fosters partnerships with companies to support them in their innovation. We acknowledge support from the Horizon Europe programme HORIZON-CL4-2022-QUANTUM-02-SGA (projects 101113690 ``PASQuanS2.1'', and 101070144 ``EURYQA''), as well as by the BPI (projects DOS0216574 ``NWQPU'' and DOS0226756 ``10XQERA'').
\end{acknowledgments}
\hfill

\bibliography{bibliography}

\end{document}